\documentclass[conference,10pt]{IEEEtran}
\IEEEoverridecommandlockouts
\usepackage{cite}
\usepackage{amsmath,amssymb,amsfonts, bm}
\usepackage{algorithmic}
\usepackage{graphicx}
\usepackage{textcomp}
\usepackage{xcolor}
\usepackage{url}
\usepackage{hyperref}
\usepackage{siunitx}

\usepackage{microtype}
\def\BibTeX{{\rm B\kern-.05em{\sc i\kern-.025em b}\kern-.08em
    T\kern-.1667em\lower.7ex\hbox{E}\kern-.125emX}}
\begin{document}

%%% for equation-related spacing
\setlength{\abovedisplayskip}{6pt}
\setlength{\belowdisplayskip}{6pt}
\setlength{\abovedisplayshortskip}{6pt}
\setlength{\belowdisplayshortskip}{6pt}
\allowdisplaybreaks[4]

\title{SHAMaNS: Sound Localization with Hybrid Alpha-Stable Spatial Measure and Neural Steerer
    \thanks{
        This work was supported by 
        JST PRESTO no. JPMJPR20CB, 
        JSPS KAKENHI nos. JP23K16912, JP23K16913, and JP24H00742 and
        ANR Project SAROUMANE (ANR-22-CE23-0011).
        All the code used to produce the results of this paper is available at \texttt{\url{https://github.com/chutlhu/shamans}}. 
        \\The first two authors contributed equally to this work.
    }
}

\author{
    \IEEEauthorblockN{
        Diego Di Carlo\IEEEauthorrefmark{1}
        \quad
        Mathieu Fontaine\IEEEauthorrefmark{3}\IEEEauthorrefmark{1}
        \quad
        Aditya Arie Nugraha\IEEEauthorrefmark{1}
        \quad
        Yoshiaki Bando\IEEEauthorrefmark{1}
        \quad
        Kazuyoshi Yoshii\IEEEauthorrefmark{4}\IEEEauthorrefmark{1}
    } \\
    \IEEEauthorblockA{
        \IEEEauthorrefmark{1}Center for Advanced Intelligence Project (AIP), RIKEN, Japan
        \\
        % \IEEEauthorrefmark{2}AIRC, National Institute of Advanced Industrial Science and Technology (AIST), Japan
        % \\
        \IEEEauthorrefmark{3}LTCI, T\'el\'ecom Paris, Institut Polytechnique de Paris, France
        \\
        \IEEEauthorrefmark{4}Graduate School of Engineering, Kyoto University, Japan
        \vspace{-1mm}
    }
}

\maketitle
\newcommand{\Hline}{\noalign{\hrule height 0.4mm}}

% text acronyms
\newcommand{\ie}{\textit{i.e.}}
\newcommand{\eg}{\textit{e.g.}}
\newcommand{\cf}{\textit{cf.}}
\newcommand{\aka}{a.k.a. }
\newcommand{\etal}{\textit{et~al.}}

% matrix operators
\newcommand{\Tr}{\mathsf{T}}
\newcommand{\Hr}{\mathsf{H}}
\newcommand{\Cr}{\mathsf{C}}

% others matrix operators
\newcommand{\trace}[1]{\text{tr}\!\left({#1}\right)}
\newcommand{\logdet}[1]{\text{logdet}\!\left({#1}\right)}
\newcommand{\Diag}[1]{\text{Diag}\!\left({#1}\right)}
\newcommand{\diag}[1]{\text{diag}\!\left({#1}\right)}

% equalities
\newcommand{\eqc}{\overset{\text{\fontsize{6pt}{0pt}\selectfont c}}{=}}

% probability laws
% \newcommand{\ComplexGaussian}[1]
% {\mathcal{N}_{\mathbb{C}}\!\left({#1}\right)}

% \newcommand{\ComplexLaw}[1]
% {\mathcal{L}_{\mathbb{C}}\!\left({#1}\right)}

\newcommand{\PositiveStable}[2]
{\mathcal{S}^{#1}_{\mathbb{R}_+}\!\left({#2}\right)}
\newcommand{\ComplexStable} [3]
{\mathcal{S}_{#1}S_{\mathbb{C}}^{#2}\!\left({#3}\right)}

\newcommand{\ComplexStudent}[2]
{\mathcal{T}^{#1}_{\mathbb{C}}\!\left({#2}\right)}
\newcommand{\ComplexGG}[2]
{\mathcal{GG}^{#1}_{\mathbb{C}}\!\left({#2}\right)}

\newcommand{\ComplexGH}[2]
{\mathcal{GH}^{#1}_{\mathbb{C}}\!\left({#2}\right)}

\newcommand{\InverseGamma}[2]{\mathcal{IG}\!\left({#1}, {#2}\right)}
\newcommand{\Exponential}[1]{\mathcal{E}\!\left({#1}\right)}
\newcommand\GeneralizedInverseGaussian[3]{\mathcal{GIG}\!\left({#1},{#2},{#3}\right)}
\newcommand\GeneralizedInverseGaussianalt[2]{\mathcal{GIG}\!\left({#1},{#2}\right)}

%bold etc.
\newcommand{\mbA}{\mathbf{A}}
\newcommand{\mbB}{\mathbf{B}}
\newcommand{\mbC}{\mathbf{C}}
\newcommand{\mbD}{\mathbf{D}}
\newcommand{\mbE}{\mathbf{E}}
\newcommand{\mbF}{\mathbf{F}}
\newcommand{\mbG}{\mathbf{G}}
\newcommand{\mbH}{\mathbf{H}}
\newcommand{\mbI}{\mathbf{I}}
\newcommand{\mbJ}{\mathbf{J}}
\newcommand{\mbK}{\mathbf{K}}
\newcommand{\mbL}{\mathbf{L}}
\newcommand{\mbM}{\mathbf{M}}
\newcommand{\mbN}{\mathbf{N}}
\newcommand{\mbO}{\mathbf{O}}
\newcommand{\mbP}{\mathbf{P}}
\newcommand{\mbQ}{\mathbf{Q}}
\newcommand{\mbR}{\mathbf{R}}
\newcommand{\mbS}{\mathbf{S}}
\newcommand{\mbT}{\mathbf{T}}
\newcommand{\mbU}{\mathbf{U}}
\newcommand{\mbV}{\mathbf{V}}
\newcommand{\mbW}{\mathbf{W}}
\newcommand{\mbX}{\mathbf{X}}
\newcommand{\mbY}{\mathbf{Y}}
\newcommand{\mbZ}{\mathbf{Z}}

\newcommand{\mba}{\mathbf{a}}
\newcommand{\mbb}{\mathbf{b}}
\newcommand{\mbc}{\mathbf{c}}
\newcommand{\mbd}{\mathbf{d}}
\newcommand{\mbe}{\mathbf{e}}
\newcommand{\mbf}{\mathbf{f}}
\newcommand{\mbg}{\mathbf{g}}
\newcommand{\mbh}{\mathbf{h}}
\newcommand{\mbi}{\mathbf{i}}
\newcommand{\mbj}{\mathbf{j}}
\newcommand{\mbk}{\mathbf{k}}
\newcommand{\mbl}{\mathbf{l}}
\newcommand{\mbm}{\mathbf{m}}
\newcommand{\mbn}{\mathbf{n}}
\newcommand{\mbo}{\mathbf{o}}
\newcommand{\mbp}{\mathbf{p}}
\newcommand{\mbq}{\mathbf{q}}
\newcommand{\mbr}{\mathbf{r}}
\newcommand{\mbs}{\mathbf{s}}
\newcommand{\mbt}{\mathbf{t}}
\newcommand{\mbu}{\mathbf{u}}
\newcommand{\mbv}{\mathbf{v}}
\newcommand{\mbw}{\mathbf{w}}
\newcommand{\mbx}{\mathbf{x}}
\newcommand{\mby}{\mathbf{y}}
\newcommand{\mbz}{\mathbf{z}}

\newcommand{\mbal}{\boldsymbol{\alpha}}
\newcommand{\mbbe}{\boldsymbol{\beta}}
\newcommand{\mbga}{\boldsymbol{\gamma}}
\newcommand{\mbde}{\boldsymbol{\delta}}
\newcommand{\mbep}{\boldsymbol{\epsilon}}
\newcommand{\mbze}{\boldsymbol{\zeta}}
\newcommand{\mbet}{\boldsymbol{\eta}}
\newcommand{\mbth}{\boldsymbol{\theta}}
\newcommand{\mbio}{\boldsymbol{\iota}}
\newcommand{\mbka}{\boldsymbol{\kappa}}
\newcommand{\mbla}{\boldsymbol{\lambda}}
\newcommand{\mbmu}{\boldsymbol{\mu}}
\newcommand{\mbnu}{\boldsymbol{\nu}}
\newcommand{\mbxi}{\boldsymbol{\xi}}
\newcommand{\mbpi}{\boldsymbol{\pi}}
\newcommand{\mbrh}{\boldsymbol{\rho}}
\newcommand{\mbsi}{\boldsymbol{\sigma}}
\newcommand{\mbta}{\boldsymbol{\tau}}
\newcommand{\mbup}{\boldsymbol{\upsilon}}
\newcommand{\mbph}{\boldsymbol{\phi}}
\newcommand{\mbps}{\boldsymbol{\psi}}
\newcommand{\mbom}{\boldsymbol{\omega}}

\newcommand{\mbGa}{\boldsymbol{\Gamma}}
\newcommand{\mbDe}{\boldsymbol{\Delta}}
\newcommand{\mbTh}{\boldsymbol{\Theta}}
\newcommand{\mbLa}{\boldsymbol{\Lambda}}
\newcommand{\mbXi}{\boldsymbol{\Xi}}
\newcommand{\mbPi}{\boldsymbol{\Pi}}
\newcommand{\mbSi}{\boldsymbol{\Sigma}}
\newcommand{\mbUp}{\boldsymbol{\Upsilon}}
\newcommand{\mbPh}{\boldsymbol{\Phi}}
\newcommand{\mbPs}{\boldsymbol{\Psi}}
\newcommand{\mbOm}{\boldsymbol{\Omega}}

% others 
\newcommand{\hx}{\hat{x}}
\newcommand{\titi}[1]{\tilde{\tilde{#1}}}
\newcommand{\bati}[1]{\bar{\tilde{#1}}}

%methods 
\newcommand{\Proposed}{SHAMaNS}
\newcommand{\MUSIC}{MUSIC}
\newcommand{\SRP}{SRP-PHAT}
\newcommand{\IWSSL}{IW-SSL}
\newcommand{\WSLL}{W-SSL}
\begin{abstract}
This paper describes a sound source localization (SSL) technique that combines an $\alpha$-stable model for the observed signal with a neural network-based approach for modeling steering vectors. Specifically, a physics-informed neural network, referred to as \textit{Neural Steerer}, is used to interpolate measured steering vectors (SVs) on a fixed microphone array.
This allows for a more robust estimation of the so-called \textit{$\alpha$-stable spatial measure}, which represents the most plausible direction of arrival (DOA) of a target signal.
As an $\alpha$-stable model for the non-Gaussian case ($\alpha \in (0,2)$) theoretically defines a unique spatial measure, we choose to leverage it to account for residual reconstruction error of the Neural Steerer in the downstream tasks. The objective scores indicate that our proposed technique outperforms state-of-the-art methods in the case of multiple sound sources.
\end{abstract}


\section{Introduction}
Sound source localization (SSL) remains a fundamental task in machine listening applications, including augmented listening, which immerses users in a coherent mixed-reality audio scene \cite{corey2019microphone, gupta2022augmented}, as well as robotics \cite{jalayer2024convlstm, wang2024slam} and autonomous driving systems \cite{madan2024acoustic, marques2022microphone}. 
These applications introduce various challenges for SSL, such as moving sensors, real-time constraints, and the need to adapt to a wide range of noisy environments.
SSL techniques can generally be categorized into three main approaches: acoustic and signal processing, data-driven deep learning, and hybrid-based signal processing.

Acoustic and signal processing techniques are widely used in SSL, such as the MUltiple SIgnal Classification (MUSIC) algorithm \cite{schmidt1986multiple}, which estimates the target signal eigenspace from the covariance matrix's eigenvectors, and the steered response power using the phase transform (SRP-PHAT) \cite{dibiase2001robust}, which uses phase correlation to estimate the time difference of arrival (TDOA). Improvements over the years, like NormMUSIC \cite{salvati2014incoherent} for frequency-axis normalization and GEVD-MUSIC \cite{nakamura2009intelligent} for enhanced noise robustness, have strengthened these methods. Modifications to SRP-PHAT, such as in \cite{cobos2010modified}, improve performance and speed by constraining the inter-microphone time delay function. However, these techniques degrade when sources are too close or there are too few microphones.

%%%%%MF comment: not very informative but let see later if it is useful. 
% Note that, in practice, all those techniques require us to know in advance the number of sources that we aim to localize. 
\begin{figure}
    \centering
    \includegraphics[width=0.9\linewidth]{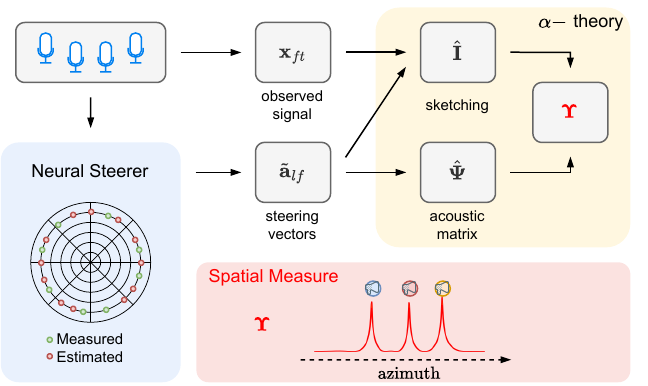}
    \caption{Schematic pipeline of the proposed algorithm.}
    \label{fig:enter-label}
\end{figure}

The SSL problem is inherently non-linear, and many SSL algorithms use deep neural networks (DNNs) to address this challenge by exploiting various audio-spatial features\cite{adavanne2021differentiable, rho2021combination, yang2021learning, subramanian2021directional}.
% While most are supervised, weakly supervised \cite{he2021neural} and semi-supervised \cite{bianco2021semi} methods also exist.
% These DNNs typically leverage features like generalized cross-correlation with phase transform \cite{adavanne2021differentiable, nguyen2021general}, Log-Mel spectrograms \cite{rho2021combination}, Log magnitude \cite{yang2021learning}, or internal phase difference \cite{subramanian2021directional}
% to predict source locations in polar or Cartesian coordinates. 
% These DNNs leverage various audio-spatial features like generalized cross-correlation with phase transform \cite{adavanne2021differentiable, nguyen2021general}, Log-Mel spectrograms \cite{rho2021combination}, Log magnitude \cite{yang2021learning}, or internal phase difference \cite{subramanian2021directional}
% to predict source locations in polar or Cartesian coordinates. 
Convolutional recurrent neural networks have been widely used for capturing SSL features, with recurrent layers benefiting from the time-series nature of audio signals. Attention-based DNNs, such as those in \cite{cao2021improved, schymura2021pilot}, focus on the most relevant features in multichannel data over time. A key challenge is that these data-driven approaches may struggle when applied to data far from the training distribution, particularly in out-of-domain scenarios.

A popular approach for balancing data-driven and fundamental model components is to combine DNN architectures, signal processing, machine learning, or stochastic models \cite{letzelter2023resilient, fontaine2017sketching, azcarreta2018permutation, duong2013spatial, fontaine2022elliptically, sumura2024joint}. The fusion of stochastic models with signal processing is common due to the stochastic component's ability to incorporate uncertainty. In \cite{schymura2021pilot}, a transformer DNN estimates an angle-of-arrival vector following a Gaussian distribution, improving performance over its non-stochastic counterpart. In \cite{sumura2024joint, asano2013sound}, the signal is modelled as a Gaussian mixture with an inverse Wishart prior on the spatial covariance matrix, improving SSL results by controlling uncertainty in steering vectors (SVs).
% Alternatively, \cite{fontaine2017sketching} models the mixture as a linear combination of SVs pointing toward $\alpha$-stable sources, generating a spatial measure heatmap. However, results using algebraic steering vectors are limited, especially without many microphones. Using room impulse response as a steering vector \cite{fontaine2017scalable} generally yields better results. However, it may not be realistic, suggesting the need for a small set of measured steering vectors and interpolation.
Alternatively, in \cite{fontaine2017sketching}, the observed mixture signal is modeled as a linear combination of $\alpha$-stable sources,
whose SVs enable localization via a spatial measure heatmap.
Although its performance was limited when the heatmap was obtained using algebraic SVs, especially with a few microphones,
it generally performed well when measured SVs obtained from recorded room impulse responses (RIRs) were used \cite{fontaine2017scalable}.
This raises the need for either a sufficient number of measured SVs or an interpolation technique to spatially upsample limited measurements.

SV interpolation is a traditional topic which includes pure signal processing interpolation, acoustics modeling and array processing. As steering vectors can be extended to head-related transfer functions (HRTFs), RIRs and directivity patterns, several interpolation techniques can be found in related HRTF upsampling~\cite{bruschi2024review}, sound field reconstruction~\cite{koyama2025physics} and array manifolds learning~\cite{manikas2004differential}. Recently, physics-informed deep learning (PIDL) approaches that combine the expressiveness of deep learning with guarantees of physical models~\cite{koyama2025physics} have shown promising results in the case of scarce measurements.

In this paper, we propose a novel fusion of the Neural Steerer \cite{di2024neural}, an existing PIDL for steering vector interpolation, with the $\alpha$-stable SSL technique \cite{fontaine2017scalable}. Our evaluation demonstrates that the proposed method, Sound Localization with Hybrid Alpha-stable Spatial Measure and Neural Steerer ({\Proposed}), outperforms other SSL baselines, particularly in scenarios with few steering vectors measured or high number of sources.

% Why SSL is important

% Trends in SSL

% Why Steering vectors are important

\section{Background}\label{sec:related_works}

% This section summarizes the background of the proposed algorithms. We overview in Section~\ref{sec:related_SSL} the $\alpha$-stable spatial measure sound source localization (SSL) technique linked to the proposed method and Section~\ref{sec:related_SVI} the one related to steering vectors interpolator. 
This section briefly describes the SSL technique based on $\alpha$-stable spatial measure (Section~\ref{sec:related_SSL})
and the SV interpolation techniques (Section~\ref{sec:related_SVI}).
We work in the short-time Fourier transform (STFT) domain,
where a time-frequency (TF) bin is identified by its frequency index $f \in [1, F]$ and time index $t \in [1, T]$ with $F$ and $T$ are the total numbers of frequency bins and time frames, respectively.

\subsection{$\alpha$-stable Spatial Measure for Sound Source Localization} \label{sec:related_SSL}

% In \cite{fontaine2017sketching, fontaine2017scalable} the multichannel observed signal $\mbx_{ft} \in \mathbb{C}^{M}$
% % in the short-time Fourier transform (STFT) domain for the time-frequency bin (TF) $f,t  \in \{1, \dots, F\} \times \{1, \dots, T\}$ where $F,T$ are the total number of frequency bins and time frame respectively
% is assumed to be decomposed as follows with $M$ the number of microphones:
% \begin{align}
% \mbx_{ft} = \sum_{l=1}^{L}\mba_{lf}s_{lft}  \label{eq:mix_model}
% \end{align}
% where $\mba_{lf} \in \mathbb{C}^M$ are the direct-path steering vector pointing toward $s_{lft}\in \mathbb{C}$. Here, $l \in [1, L]$ represents one potential position on a grid (can be azimuth elevations or cartesian coordinates for instance) of a potential active source. We denote $N <  L$ being the number of sources we aim to localize.

Following \cite{fontaine2017sketching, fontaine2017scalable},
an $M$-channel observed mixture signal $\mbx_{ft} \in \mathbb{C}^{M}$
is assumed to be a sum of source images propagated from $L$ potential source locations:
\begin{align}
\mbx_{ft} = \sum_{l=1}^{L}\mba_{lf}s_{lft} , \label{eq:mix_model}
\end{align}
where $\mba_{lf} \in \mathbb{C}^M$ is the direct-path SV pointing toward $s_{lft}\in \mathbb{C}$
and $l \in [1, L]$ represents one potential source location on a grid (e.g., in the polar or Cartesian coordinate system).
Given $\mbx_{ft}$ and the number of sources of interest $N$, we aim to estimate the locations of $N$ sources ($N < L$).

% The assumption made here is that the $N$ sources of interest can be seen as ``outliers'' in terms of energy in their respective position and that the rest can be considered as noise. 

Each \( s_{lft} \) is assumed to follow a complex univariate isotropic \( \alpha \)-stable distribution \cite{samoradnitsky1994stable} with a frequency- and time-invariant scale parameter $\Upsilon_l \in \mathbb{R}_+$, denoted as \( s_{lft} \sim \ComplexStable{\alpha}{}{\Upsilon_l} \).
The $N$ active source signals coming from their respective locations can be seen as ``outliers'' in terms of energy since $\Upsilon_{l^\prime} \approx 0$ when $l^\prime$ is not a position of an active source.
% Here, \( \ComplexStable{\alpha}{}{\sigma} \) represents a univariate complex isotropic \( \alpha \)-stable distribution with a scale parameter \( \sigma \geq 0 \) and a characteristic exponent \( \alpha \in (0,2) \).
The characteristic exponent $\alpha \in (0,2]$ determines the heaviness of the distribution's tail: smaller \( \alpha \) leads to heavier tails (which allows more outliers),
whereas \( \alpha = 2 \) corresponds to the Gaussian distribution.
For the non-Gaussian case ($\alpha < 2$), it can be shown that $\mbx_{ft}$ follows a complex isotropic $\alpha$-stable distribution with a unique discrete \textit{spatial measure} $\mbUp \triangleq [\Upsilon_1, \dots, \Upsilon_L] \in \mathbb{R}_+^{L}$ \cite{fontaine2017scalable}.

Due to the absence of an analytic form for the probability density function,
we use
the time-invariant Lévy exponent $I_f(\mbth) \in \mathbb{R}$ given a parameter $\mbth \in \mathbb{C}^{M}$
to establish a relationship between $\mbUp$, $\mbx_{ft}$, and
% a normalized SV $\tilde{\mba}_{lf} \triangleq \mba_{lf} / \Vert \mba_{lf} \Vert_2^2$
$\mba_{lf}$
as
\begin{align}
I_{f}(\mbth) \triangleq  -\ln \mathbb{E}[e^{\mathrm{i}\Re(\mbth^{\mathsf{H}})\mbx_{ft}}] = \sum_{l=1}^L \Upsilon_l\left|\mbth^\mathsf{H}{\tilde{\mba}}_{lf}\right|^\alpha ,
\label{eq:levy_exp}
\end{align}
where $\tilde{\mba}_{lf} \triangleq \mba_{lf} / \Vert \mba_{lf} \Vert_2^2$ is a normalized SV.

Following \cite{fontaine2017scalable},
instead of using a Lévy exponent given a single parameter $\mbth$ as above,
we consider a set of parameters $\mbth_{lf} \triangleq \tilde{\mba}_{lf}$ to provide physical interpretation
and use the nonnegative estimator of the Lévy exponent given by
\begin{align}
\hat{I}_f(\tilde{\mba}_{lf}) = -2\ln \left| \frac{1}{T} \sum_{t=1}^{T} \exp\left(\mathrm{i}\frac{\Re(\tilde{\mba}_{lf}^\mathsf{H}\mbx_{ft})}{2^{1/\alpha}}\right)\right| .
\end{align}
By defining $\hat{\mbI}_f \triangleq [\hat{I}_f(\tilde{\mba}_{1f}), \dots, \hat{I}_f(\tilde{\mba}_{Lf})]^\top \in \mathbb{R}_+^{L}$  
and $\mbPs_f \triangleq \big[\big|\tilde{\mba}_{lf}^\mathsf{H}\tilde{\mba}_{l^\prime f}\big|^\alpha\big]_{l=1,l^\prime = 1}^{L,L} \in \mathbb{R}_+^{L \times L}$,
we have the relation $\hat{\mbI}_f \approx \mbPs_f\mbUp$.
% It was then shown in \cite{fontaine2017scalable} that a nonnegative estimator of $I_f$ can be formulated. In particular, an empirical approach that considers a set $\mbth_{lf}$ equal to the normalized steering vectors $\tilde{\mba}_{lf}$ appears to be a good trade-off, providing an interpretable set of samples derived from the physical properties of the model.
% If we define $\hat{\mbI}_f \triangleq [\hat{I}_f(\tilde{\mba}_{1f}), \dots, \hat{I}_f(\tilde{\mba}_{Lf})]^\top \in \mathbb{R}_+^{L}$, where  
% \begin{align}
% \hat{I}_f(\tilde{\mba}_{lf}) = -2\ln \left| \frac{1}{T} \sum_{t=1}^{T} \exp\left(\mathrm{i}\frac{\Re(\tilde{\mba}_{lf}^\mathsf{H}\mbx_{ft})}{2^{1/\alpha}}\right)\right|
% \end{align}
% and define $\mbPs_f \triangleq \left[\left|\tilde{\mba}_{lf}^\mathsf{H}\tilde{\mba}_{l^\prime f}\right|^\alpha\right]_{l=1,l^\prime = 1}^{L,L} \in \mathbb{R}_+^{L \times L}$, we obtain the approximation $\mbI_f \approx \mbPs_f\mbUp$.
Since $\mbUp$ is frequency-independent, we concatenate all $\hat{\mbI}_f$ and $\mbPs_f$ matrices along the frequency axis to form a vector $\hat{\mbI} \in \mathbb{R}_+^{FL}$ and a matrix $\mbPs \in \mathbb{R}_+^{FL \times L}$, respectively, leading to the relation:
\begin{align}
\hat{\mbI} \approx \mbPs\mbUp. \label{eq:approx_I}
\end{align}
As $\mbPs$ may still be ill-conditioned making it inversion unstable,
we instead minimize the pointwise $\beta$-divergence%
\footnote{It corresponds to the Itakura-Saito divergence ($\beta=0$), the Kullback-Leibler divergence ($\beta=1$), or the Euclidean distance ($\beta=2$)}
between the left- and right-hand sides of Eq.~\eqref{eq:approx_I},
% we aim to minimize the discrepancy between the left- and right-hand sides of Eq.~\eqref{eq:approx_I}.
% To achieve this, %
% denoted $d_\beta (\cdot | \cdot)$ (e.g., for $\beta=0, \beta=1, \beta=2$, this corresponds to the Itakura-Saito, Kullback-Leibler, and Euclidean divergences, respectively),
% while imposing a penalty constraint on $\mbUp$ due to its high sparsity \cite{fontaine2017sketching}.
while imposing a sparsity penalty constraint on $\mbUp$ \cite{fontaine2017sketching}.
This optimization problem ultimately leads to the following iterative update strategy:  
\begin{align}
\hat{\mbUp} \leftarrow \hat{\mbUp} \odot \frac{\mbPs^\top\left(\left(\mbPs\hat{\mbUp}\right)^{\beta-2} \odot \hat{\mbI}\right)}{\mbPs^\top\left(\left(\mbPs\hat{\mbUp}\right)^{\beta-1}\right) + \lambda} , \label{eq:updates}
\end{align}  
where $\lambda \in \mathbb{R}_+$ is the sparsity penalty coefficient
with $a \odot b$ and $\frac{a}{b}$ denote element-wise multiplication and division, respectively.  

% % Experimental results in \cite{fontaine2017scalable, fontaine2017sketching} demonstrate that 
% This approach performed well with many microphones in a reverberant environment \cite{fontaine2017scalable, fontaine2017sketching}.
% However, a performance drop was observed when algebraic SVs, instead of measured room impulse responses (RIRs), were used as $\tilde{\mba}_{lf}$.
% In this paper, we propose to
% extend the model in Eq.~\eqref{eq:mix_model} %by adding noise term
% to be theoretically applicable in noisy real-world environments
% and use interpolated measured SVs
% to cope with the costly SV measurement issue.

\subsection{Steering Vector Interpolation}\label{sec:related_SVI}

SVs have been typically used 
 for representing the spatial characteristics 
 of the sound field around an array aperture as function of the probing direction. 
Given the microphone array geometry and a target location,
the \textit{algebraic} model for the SV
% $a_{mf}$ for the $m$-th microphone at frequency index $f$ reads
is given by
\begin{equation}
    \mathbf{a}_{lf} = \exp(2 \pi \omega_f \mathbf{r}_{l} / c F_s) / \sqrt{4\pi} \mathbf{r}_{l},
\end{equation}
where $\textbf{r}_{l}$ contains the distances between the target location $l$ and the microphone positions in the array, $c$ is the speed of sound, $\omega_f$ is the angular frequency in the band $f$, and $F_s$ is the sampling frequency.

In real-world environments, frequency-dependent filtering effects, e.g., scattering around the array and pick-up patterns, alter the theoretical shape of the SVs.
% To account for these effects, measured steering vectors can be accurately measured in dedicated facilities \cite{donley2021easycom}. Alas, measuring them at high spatial resolution is a cumbersome, time-consuming task, if not unfeasible and prohibitive due to the cost and setup complexity. 
Although they can be measured accurately, as in \cite{donley2021easycom}, measuring them at high spatial resolution may be unfeasible and prohibitive due to the cost and setup complexity. 
To address this, reliable data-driven interpolation approaches have received significant attention.

For far field applications, it is common to interpolate SVs with respect to location on the sphere centered at the array location. Methods like barycentric interpolation and spherical splines provide a smooth interpolation on the spherical surface (See~\cite{bruschi2024review} for a review in case of HRTFs data).
Physically motivated methods, such as spherical harmonics (SH) interpolation, are currently the state of the art for such task provided a set of measurements on a predefined grid. The SH expansion of the $m$-th element of SV for the direction of arrival (DOA) $\vartheta_l$ is written as
\begin{equation}
    a_{lmf} = \sum_{\nu=0}^{\infty} \sum_{\mu=-\nu}^{\nu} c_{\nu\mu,mf} Y_{\nu}^\mu(\vartheta_l),
\end{equation}
where $Y_{\nu}^\mu$ is an SH basis of order $\mu$ and degree $\nu$.
Given a set of measurements at different  $\vartheta_l$, the expansion coefficients $c_{\nu\mu,mf}$ are typically estimated by regularized linear regression for each microphone $m$ and frequency band $f$ independently.
Unfortunately, these methods lead to poor interpolation when measurements are sparse and randomly distributed~\cite{bau2022estimation}.

Recent advancement in physics-informed deep learning architectures provides a new solution to increase the spatial resolution of acoustic measurements~\cite{chen2023sound,koyama2025physics}. Our previous study \cite{di2024neural} demonstrated that a coordinate-based neural network, called Neural Steerer (NS), can be used to up-sample steering vectors from a few sparse measurements.

\section{Proposed Extension}

This section introduces the proposed Sound localization with Hybrid Alpha-stable Spatial Measure and Neural Steerer ({\Proposed}).

\subsection{Mixing Model with Additive Noise}

% In Eq.~\eqref{eq:mix_model}, we notice that no noise is included. In this section, we aim to first add such additive noise and show that, under some assumptions, adding a noise component will not change the estimation procedure. 
% We replace Eq.~\eqref{eq:mix_model} with:

% To be theoretically applicable in noisy real-world environments,
We formulate a new mixing model by introducing an additive noise component $\mbn_{ft}$ to the model in Eq.~\eqref{eq:mix_model} as follows:
\begin{align}
    \mbx_{ft} = \sum_{l=1}^{L}\mba_{lf}s_{lft} + \mbn_{ft}  \label{eq:new_mix_model} ,
\end{align}
where
% the probabilistic model remains the same for the scalar source, and 
$\mbn_{ft} \sim \mathcal{E}_\alpha(\epsilon \mathbb{I}_M)$ follows a complex isotropic elliptically contoured $\alpha$-stable distribution
% independent of each $s_{lft}$ and for all $f,t$, 
with $\epsilon \in \mathbb{R}_+$ and $\mathbb{I}_M \in \mathbb{R}^{M\times M}$ is the identity matrix \cite{fontaine2022elliptically}. 
% It can be easily shown that $I_f$ is in this case $\forall \mbth \in \mathbb{C}^M$:
The Lévy exponent $I_f(\mbth) \in \mathbb{R}$ given $\mbth \in \mathbb{C}^{M}$ is then expressed as
\begin{align}
I_f(\mbth) = \sum_{l=1}^L \Upsilon_l\left|\mbth^\mathsf{H}{\tilde{\mba}}_{lf}\right|^\alpha + C_\alpha \left|\mbth^\mathsf{H}\mbth\right|^\alpha \label{eq:new_Levy} ,
\end{align}
where $C_\alpha = (\epsilon/2)^{\alpha/2}$. 
If we consider a set of $\mbth_{lf} \triangleq \tilde{\mba}_{lf}$ to compute $\mbPs_f$, $\mbUp$, $\mbI_f$ as in Section~\ref{sec:related_SSL}, we obtain the relation:
% If we consequently evaluate on $\mbth = \tilde{\mba}_{lf}$ and use $\mbPs_f, \mbUp, \mbI_f$ as defined in Section~\ref{sec:related_SSL}, then we get from Eq.~\eqref{eq:new_Levy}:
\begin{align}
 \mbI_f = \mbPs_f\mbUp + C_\alpha\bold{1}_L ,  \label{eq:new_Levy2}
\end{align}
where $\bold{1}_L$ is an $L$-dimensional vector of ones.
Interestingly, deriving the multiplicative update by computing the gradient along $\mbUp$ makes the second term in Eq.~$\eqref{eq:new_Levy2}$ vanish and results in the same update formula as in Eq.~\eqref{eq:updates}.
% If we then derive the multiplicative update that is just based on computing the gradient along $\mbUp$, then the second term in $\eqref{eq:new_Levy2}$ vanishes, and we get the same update as in Eq.~\eqref{eq:updates}. 

This result implies that if $\alpha$ for $\mbx_{ft}$ is correctly estimated, the noise component is implicitly taken care of and the proposed approach can extract only the desired signals.
From this viewpoint, we claim that a correct estimation of $\alpha$, rather than a cherry-picking approach \cite{fontaine2017scalable, fontaine2017sketching}, is essential for the algorithm to work correctly. 
Note that %it is not applicable in the Gaussian case since 
the spatial measure of $\mbn_{ft}$ is not unique in the Gaussian case
% in this case 
so estimating the spatial measure from $\mbx_{ft}$ 
is theoretically incorrect in this case \cite{samoradnitsky1994stable}. 

\subsection{Observation Normalization for Increasing Robustness}

% Another improvement that we propose in our paper is to consider a "normalized version" of $\mbx_{ft}$ to compute Eq.~\eqref{eq:new_Levy2}.
We propose to normalize $\mbx_{ft}$ for computing Eq.~\eqref{eq:new_Levy2}.
% Intuitively, 
$\alpha$-stable realizations for $\alpha < 2$ would have outliers that benefit SSL,
% Intuitively, $\alpha$-stable realizations for $\alpha < 2$ represent outliers that benefit SSL, as the largest outliers will be identified among all potential candidates on the grid.
but a sound source with low energy may not be correctly identified, leading to poor SSL results in practice.
% Moreover, the characteristic frequencies $\mbth_{lf} = \tilde{\mba}_{lf}$ used to measure the Lévy exponent belong to the hypersphere.
% It can be shown using 
According to the covariation formula \cite[p. 87]{samoradnitsky1994stable},
the spatial measure remains computationally unchanged up to a scale parameter independent of the direction.
We consequently choose to use $\tilde{\mbx}_{ft} \triangleq \mbx_{ft} / \Vert \mbx_{ft} \Vert_p^p$, where $p < \alpha$ since the norm can be theoretically infinite for $p\geq\alpha$.
% Empirical studies reveal a considerable difference between the normalized and non-normalized versions, leading us to retain the normalized extension in our experiments.

\subsection{Fusion with Neural Steerer}

Ideally, we can interchangeably use algebraic SVs or measured SVs as $\mbth_{lf}$ to compute $\hat{\mbI}$ and $\mbPs$.
However, algebraic SVs significantly differ from SVs measured in the real world, resulting in poor SSL performance in practice.
Since measuring SVs is costly, this paper proposes to generate a large set of SVs from a small set of measured SVs.
% by interpolation using Neural Steerer
To this end, we modify the Neural Steerer (NS) model~\cite{di2024neural}. Instead of learning the mapping from a DOA to its steering vector filter, we use a coordinate-based neural network to estimate the SH expansion coefficients $c_{\nu\mu,mf}$ for a given microphone index, DOA Cartesian coordinate and frequency band.\footnote{Details can be found in the supplementary materials in the code website.} 
The expressiveness of the deep learning helps to resolve the higher SH coefficients. The extensive study of the influence of different interpolation methods on SSL performances is out of the scope of this work.

% the errors are roughly localized around the same frequency regardless of direction.
We assume that we have $\mba_{lf}^{\text{NS}} \approx \mba_{lf}^{\text{true}} + \mbe_{f}$ where $\mba_{lf}^{\text{NS}}, \mba_{lf}^{\text{true}}, \mbe_{f}$ denote the NS-reconstructed SV, the true SV, and the error, respectively. 
As is common for interpolation methods of acoustic measurements, reconstruction error increases with the frequencies~\cite{koyama2025physics}.
Consequently, error appears impulsive at certain frequencies and directions, thus, we can assume that it is incorporated into the noise component $\mbn_{ft}$. 

\section{Evaluation}
% We test our hypotheses through computer simulations. In the following, we describe the simulation setup, the methods and we discuss the results.

This section presents the SSL performance of {\Proposed} in comparison to two widely-used wideband SSL methods: MUSIC~\cite{schmidt1986multiple} and SRP-PHAT~\cite{dibiase2001robust}.
Each method may use the measured oracle SVs (``Ref.''), the corresponding algebraic SVs (``Alg.''), or the interpolated SVs generated by NS.
The SSL performances were evaluated in terms of
the \textit{angular error} $d(\vartheta,\hat{\vartheta})$ on the unit circle and 
the \textit{accuracy} for $d(\vartheta,\hat{\vartheta}) < 15^\circ$.

\subsection{Settings}

We used
the SVs measured using head-worn smartglasses, featuring four microphones on the frame and binaural in-ear microphones ($M = 6$),
from the SPeech Enhancement for Augmented Reality (SPEAR) challenge dataset \cite{guiraud22spear}.
% which was derived from the EasyCom dataset \cite{donley2021easycom},
From these $1020$ anechoic SVs measured using a dummy head, %in an anechoic condition,
we randomly sampled $N_\text{SV} = [8, 16, 32, 64, 128]$ measurements on the entire sphere to fit an NS model (``NS-$N_\text{SV}$'').
% We compared the SSL performance against the reference ground truth measurements (Ref.) and the corresponding algebraic SVs (Alg.).

We generated 30 different acoustic scenes for each investigation below by varying the number of sources $N$, the signal-to-noise ratio (SNR) for additive white Gaussian noise, and reverberation time (RT60). For each acoustic scene, we generated RIRs and convolved them with speech utterances from the VCTK corpus (ver. 0.92)~\cite{yamagishi2019cstr}. To obtain directional responses, each RIR was spatially convolved with the corresponding SV which is interpolated using SH interpolation from the oracle measured SVs.
The microphone array was placed randomly in a shoebox room with a size of $6 \times 4 \times 3$ \si{\metre}.
% Sound sources were placed at different locations on the azimuthal plane (elevation = 0) at a distance of $1.7$ meter.
The potential source locations were $6$ degrees apart ($L=60$) on the azimuthal plane (elevation $= 0$) at a distance of $1.7$ \si{\metre} from the array.
The sources were then randomly placed at least $12$ degrees apart. 

All data were sampled at $F_s=48$ kHz. The STFT frame size was 768 samples with 50\% overlap between frames and computed with a Hann window. Otherwise specified, only positive frequencies up to 8 kHz were considered.

% The SSL performances were computed in terms of angular error $d(\vartheta,\hat{\vartheta})$ on the unit circle and accuracy defined as $d(\vartheta,\hat{\vartheta}) < 15^\circ$.

% The proposed methods 
For {\Proposed}, we estimated $\alpha$ from the mixture as in \cite{fontaine2021alpha}, fixed $p=1$, and then optimized the parameters using Eq.~\eqref{eq:updates}
with %$L=60$ (one DOA every 6 degrees),
$\beta=1$ and $\lambda=10^{-3}$ for 500 iterations.
The estimated spatial measure was initialized as $\hat{\mbUp} \leftarrow \mathbf{1}_{L}$. %and $\alpha$ is estimated using the setting and method described in \cite{fontaine2021alpha}.
% We compared the performance of the proposed method {\Proposed} to that of other popular wideband SSL algorithms: MUSIC~\cite{schmidt1986multiple} and SRP-PHAT~\cite{dibiase2001robust}.
% We use the notation MUSIC-$N$ in this work to highlight the number of sources used for the method.
For MUSIC, we used either
the only largest eigenvector (``MUSIC-1'') or the first four largest eigenvectors (``MUSIC-4'')
to define the signal subspace.
In the case of $N>1$, the Hungarian algorithm was used to find the best-estimated source permutation.

 \begin{figure}[t]
     \centering
     \includegraphics[width=\linewidth]{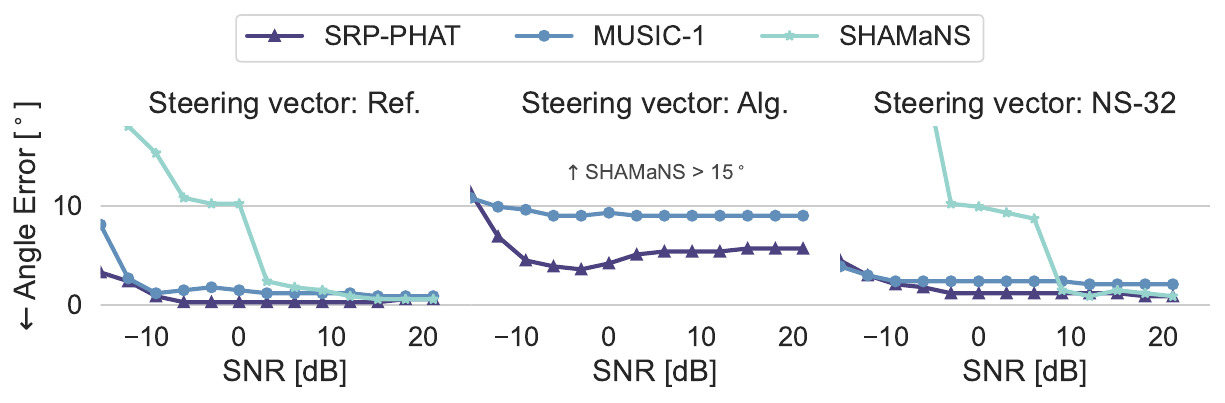}
     \\\vspace{0.3em}
     \includegraphics[width=\linewidth]{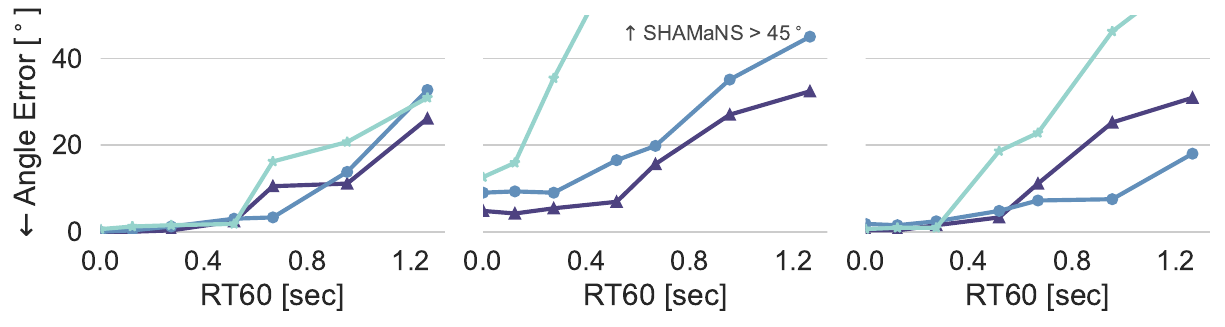}
     \caption{Average angular errors for the different SNRs (top) and RT60s (bottom) using the different localization methods and steering vector models.}
     \label{fig:snr-rt60}
 \end{figure}

\begin{figure}[t]
    \centering
    \includegraphics[width=\linewidth]{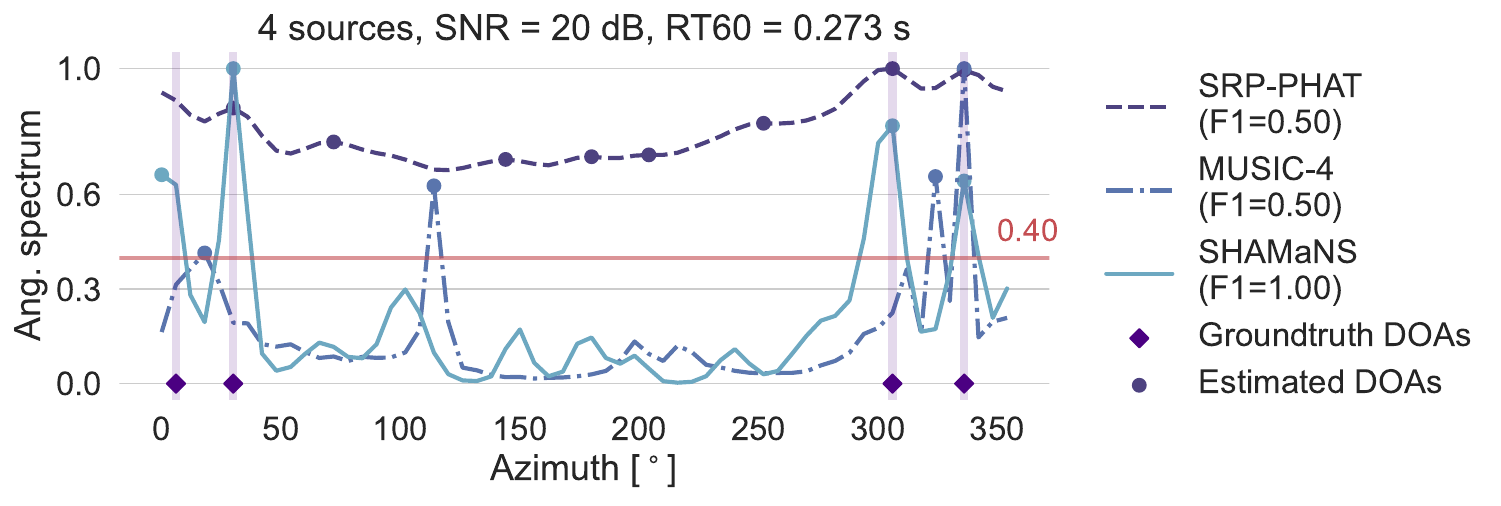}
    \caption{SSL cost function (angular spectrum) for different localization methods. Horizontal lines represent the threshold for peak-picking methods.}
    \label{fig:ang-spec}
\end{figure}

\subsection{Results}

Fig.~\ref{fig:snr-rt60} reports the average angular error against different noise and reverberation levels in a single sound source case ($N=1$)
with 30 random %Monte Carlo
simulations for each value of SNR and RT60.
{\Proposed} was shown to have comparable performances for positive SNR and moderate reverberation when measured or interpolated SVs were used, leading to an average error of $0.98^\circ \pm 2.45^\circ$ for {\Proposed}, $1.07^\circ \pm 2.31^\circ$ for MUSIC-1, and $0.38^\circ \pm 1.47^\circ$ for SRP-PHAT.
The performances of all methods dropped when the algebraic SVs were used, where SRP-PHAT outperformed with an average error of 
$5.48^\circ \pm 7.50^\circ$.

\begin{figure}[t]
    \centering
    \includegraphics[width=\linewidth]{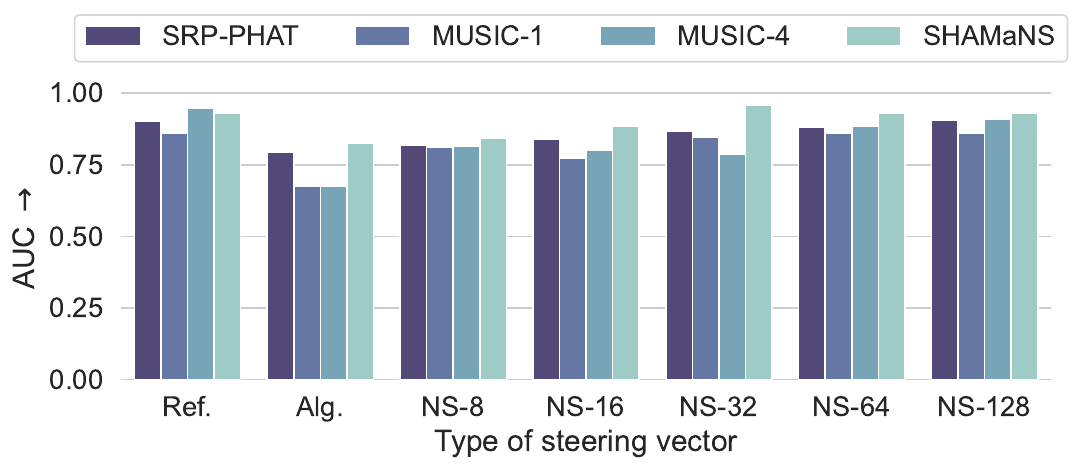}
    \vspace{-\baselineskip}
    \caption{Area under the ROC curve for the number of source classifications using different thresholds of the angular spectrum. Comparison is for different localization methods using different steering vector models.}
    \label{fig:auc}
\end{figure}

\begin{figure}[t]
    \centering
    \includegraphics[width=\linewidth]{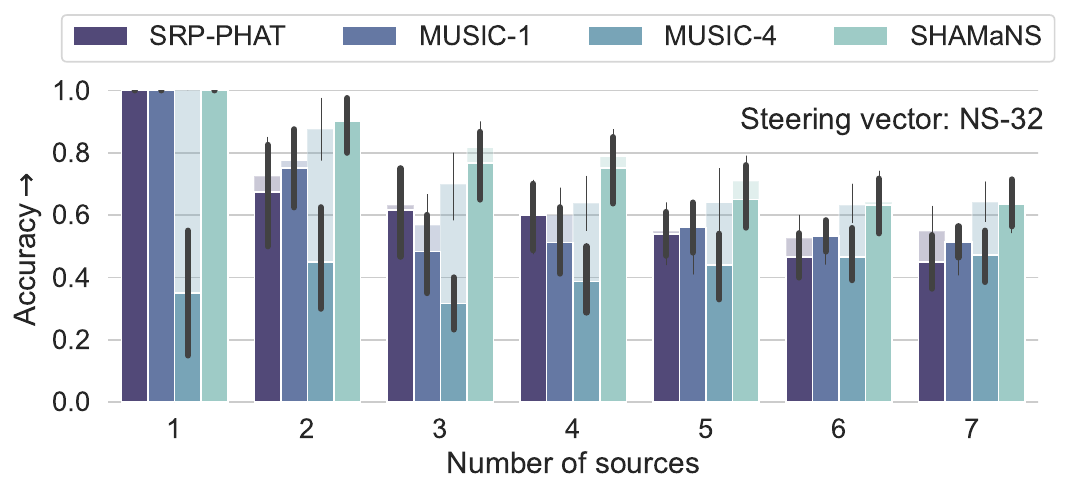}
    \vspace{-\baselineskip}
    \caption{Accuracy (success for angular error less than 15 degrees) for different localization methods and steering vector models. RT60=0.123 ms, SNR = 20 dB. Performances with SV = Ref. are reported in the background in transparency.}
    \label{fig:num-src}
\end{figure}

Next, we investigated the impact of the SV interpolation quality on the SSL performance
% how the performances vary with the quality of the SV interpolation method 
when the number of sources $N$ is unknown.
This task was achieved %can be complied with 
by thresholding the normalized SSL method's cost function for the candidate DOAs (shown in Fig~\ref{fig:ang-spec}) and counting the highest peaks. By considering it as a classification task, we then computed the area under the curve (AUC) using different value thresholds to perform the classification.
% Here we considered two variants of the MUSIC methods, for $N=1$ (MUSIC-1) and $N=4$ (MUSIC-4), using the first and first four subspace eigenvectors, respectively.
The performance metrics were computed
% from 30 random simulations 
setting $N=3$, SNR $=20$ dB and RT60 $=0.273$ ms.
The results in Fig.~\ref{fig:auc} show that {\Proposed} and MUSIC-4 outperformed the other baselines. As expected, using interpolation methods that used more observations improves the performance. The oracle SSL performance could be obtained with only 10\% of the initial measurements and a reasonable performance with only 32 random measurements.

Finally, we studied the accuracy of locating $N$ sound sources as a function of their numbers.
The results in Fig.~\ref{fig:num-src} show that most of the methods could estimate source location when $N<5$ with accuracy above 55\% when either oracle or sufficiently well-interpolated SVs (NS-32). Performances dropped for higher values of $N$, while {\Proposed} outperformed the baseline methods for locating from 2 to 6 concurrent sources, which happened in most practical use cases.

\section{Conclusion \& Future Work}
This study introduced a novel approach that combines NeuralSteerer, a deep neural network-based physics-informed steering vector model, with a theoretically grounded source localization technique derived from $\alpha$-stable theory. 
We provide a theoretical justification for this hybrid fusion by incorporating the error model of NeuralSteerer into the additive impulsive noise framework of the $\alpha$-stable model. 
Our results demonstrate that the proposed method achieves superior performance for localizing multiple sound sources, even when only a limited number of measured steering vectors are available.
Future work will study the interaction between different interpolation and localization methods.

\bibliographystyle{IEEEtran}
\bibliography{refs}

\end{document}